\documentclass[twocolumn,showpacs,preprintnumbers,superscriptaddress,
amsmath,amssymb]{revtex4}

\usepackage{epsfig}
\usepackage{graphicx}
\usepackage{dcolumn}
\usepackage{bm}
\usepackage{times}
\usepackage{xcolor}

\begin{document}


\title{Epidemic spreading on complex networks with general degree and weight distributions}

\author{Wei Wang}
\affiliation{Web Sciences Center, University of Electronic
Science and Technology of China, Chengdu 610054, China}

\author{Ming Tang}\email{tangminghuang521@hotmail.com}
\affiliation{Web Sciences Center, University of Electronic
Science and Technology of China, Chengdu 610054, China}
\affiliation{Center for Atmospheric Remote Sensing(CARE),
Kyungpook National University, Daegu, 702-701, South Korea}

\author{Hai-Feng Zhang}
\affiliation{
School of Mathematical Science, Anhui University, Hefei 230039, China
}

\author{Hui Gao}
\affiliation{Web Sciences Center, University of Electronic
Science and Technology of China, Chengdu 610054, China}

\author{Younghae Do}
\affiliation{Department of Mathematics, Kyungpook National
University, Daegu 702-701, South Korea}

\author{Zong-Hua Liu}
\affiliation{Department of Physics, East China Normal University,
Shanghai 200062, China}

\date{\today}

\begin{abstract}
The spread of disease on complex networks has attracted
widely attention in the physics community. Recent works have demonstrated that
heterogeneous degree and weight distributions have a significant influence on the
epidemic dynamics. In this study, a novel edge-weight based compartmental
approach is developed to estimate the epidemic threshold and epidemic size (final infected density)
on networks with general degree and weight distributions, and a remarkable agreement
with numerics is obtained. Even in complex network with the strong heterogeneous degree and weight
distributions, this approach is worked.
We then propose an edge-weight based removal strategy with different biases,
and find that such a strategy can effectively control the spread
of epidemic when the highly weighted edges are preferentially removed,
especially when the weight distribution of a network is extremely heterogenous.
The theoretical results from the suggested method
can accurately predict the above removal effectiveness.

\end{abstract}

\maketitle

\section{Introduction} \label{sec:intro}

In most real-world networks, edges connecting two nodes are
often associated with weights that differentiate them in terms of
their strength, intensity, or capacity~\cite{Http}. For example, in
 scientific collaboration networks, the weight of each edge can stand for
the number of papers that two authors have coauthored~\cite{Newman2001,Barrat2004};
in communication networks, it can represent the total duration of calls
between two users over a period of time~\cite{Onnela2007};
and in brain networks, it can be viewed as
the times of memories reinforced between neurons~\cite{Zhuo2011,Amaral2000};
also, it can account for the number of passengers between
two airports in aviation networks~\cite{daRocha2009}.

A large number of empirical studies have verified that the degree and weight
distributions of many weighted networks are greatly heterogeneous~\cite{Albert:2002} (\emph{e.g.},
log-normal~\cite{Newman2001} and power-law~\cite{Barrat2004}) and
these inhomogeneous structures have remarkable effects on the
dynamical processes on the substrate of networks~\cite{Dorogovtsev2008,Holme2002}, especially for the
dynamics of epidemics~\cite{Newman2003a,Wang2014a,Gross2008,Holme2012}. Scores of researchers have proven that
the strong heterogeneity of degree distribution can reduce or even vanish
the epidemic threshold under some certain conditions [\emph{e.g.}, on the
scale-free networks of degree distribution $p(k)\sim k^{-\gamma_D}$
with degree exponent $\gamma_D\leq3$ in thermodynamic limit]~\cite{Pastor-Satorras2001,Boguna2013}.
On weighted networks, some researches have shown that the inhomogeneity of
weight distribution can also significantly affect the epidemic dynamics,
such as the epidemic threshold and epidemic prevalence~\cite{Yan2005, Yang2012,
Chu2009,Min2013,Deijfen2011,Kamp2013,Rattana2013,Sun2014}.
For instance, Zhou \emph{et al.} suggested that increasing the dispersion of weight
distribution can reduce the velocity of epidemic spreading as well
as the epidemic prevalence~\cite{Yan2005,Yang2012}.

Moreover, these heterogeneous structural properties have triggered
the improvement of immunization strategy for complex networks.
A few effective strategies have been proposed
for the networks of heterogeneous degree distributions,
including the targeted immunization strategy~\cite{Gomez-Gardenes2006,Zhang2009},
acquaintance immunization strategy~\cite{Cohen2003} and
information based immunization method~\cite{Granell2013}.
For weighted networks, Deijfen proposed a variation
of the so called acquaintance immunization strategy,
where nodes are chosen randomly and these random nodes' neighbors
with high edge-weights are vaccinated, and the modified strategy
is more effective than the classical acquaintance immunization strategy
where the neighbors are vaccinated randomly for a given vaccination coverage~\cite{Deijfen2011}.
In addition, the targeted immunization strategy based on node's strength
showed an effective immune effectiveness~\cite{Eames2009}.

Most of the existing works studying epidemic dynamics and its immunization strategy
on weighted networks have been analyzed through heterogeneous mean-field
theory (HMF)~\cite{Sun2014,Buono2013}, percolation theory~\cite{Deijfen2011} or
pairwise approximation method (PA)~\cite{Kamp2013,Rattana2013}.
The HMF theory assumes that the nodes of the same degrees will show the same dynamical characteristics
~\cite{Pastor-Satorras2001,Moreno2002,Castellano2006},
and can only qualitatively understand the effects of heterogeneous structural properties
on quenched networks~\cite{Sun2014,Buono2013}. Similar to the HMF theory,
the analytical results derived from the percolation theory will also obviously
deviate from the numerical results in the case of strong structural heterogeneity~\cite{Deijfen2011},
which is caused by the strong dynamic correlations between two connected nodes~\cite{Ferreira2012}.
The PA method can partly reflect the dynamic correlations and thus get a more accurate
theoretical prediction~\cite{Eames2002,Gross2006}.
In the PA method, a number of $E\propto O(k_{max}^{2}w_{max}^{2})$
equations is needed to govern the dynamical system, with $k_{max}$
and $w_{max}$ be the maximum degree and weight, respectively~\cite{Kamp2013,Rattana2013}.
So it will take a large amount of time to solve the nonlinear equations for epidemic dynamics
when $k_{max}$ and $w_{max}$ are very large (\emph{i.e.},
networks with strong heterogenous degree and weight distributions),
greatly limiting its ability of real-time prediction.
Yang \emph{et al.} developed an edge-based mean-field approximation
to study epidemic spreading on homogeneous networks with heterogeneous
weight distribution, but this method is not able to provide a very
accurate prediction on reality weighted networks with strong structural heterogeneity~\cite{Yang2012}.
Therefore, it is imperative for us to built
a comprehensive method to depict the spreading dynamics on networks
with general degree and weight distributions.

In this paper, we develop an edge-weight based compartmental
approach to study epidemic spreading on
networks with general degree and weight distributions. Our
theory predicts that the epidemic threshold and epidemic
size are closely related to the degree and weight
distributions,  which are in good agreement
with the results from numerical simulations.
In general, increasing the heterogeneity of weight distribution can suppress the
epidemic spreading. However, for the degree distribution,
increasing its heterogeneity can enhance (reduce) the epidemic
size at the small (large) value of unit infection probability.
We then propose an edge-weight based removal strategy to restrain
the spreading of epidemic on weighted networks.
Both the theoretical predications and experimental simulations indicate that
an epidemic can be well controlled if the edges with high weights are
preferentially removed, especially for the networks with the strong
heterogenous weight distribution and homogenous degree distribution
near the epidemic threshold.

The paper is organized as follows. In Sec. II, we describe
weighted networks with general degree and weight distributions
and the dynamical processes on it. In Sec. III, we will present an
edge-weight based compartmental approach for
the epidemic spreading and edge-weight based removal strategy on weighted networks.
Numerical confirmation of the theoretical predictions will be provided in
Sec. IV. We will draw our conclusions in Sec. V.

\section{Model} \label{sec:model}
We consider a population of size $N$ with degree distribution
$p(k)$ and weight distribution $g(w)$. For the sake of simplicity,
we assume that there is no correlation between the degree and weight
distributions (\emph{i.e.}, edge weight is independent of node's
degree). To construct a weighted network with the above degree and
weight distributions, we first built an unweighted complex network
as follows: i) generate a degree sequence following the degree distribution $p(k)$;
ii) assign a total number of $k_i$ edge stubs to each node $i$; iii) randomly select two
stubs to create an edge; iv) repeat the process iii) until there are
no stubs left. Self-loops and multiple edges between the same
pair of vertices are prohibited~\cite{Newman2010}. After that,
each edge in the unweighted network is assigned a weight
according to the weight distribution $g(w)$.
The networks generated according to the above steps have
no degree-degree and degree-weight correlations in the thermodynamic limit.

The epidemic spreading on weighted networks is described as
a weighted Susceptible-Infected-Recovery (SIR) epidemiological model.
In the SIR model, each node can be in one of the three states: susceptible state (S),
infected state (I), and recovery state (R).
To initiate an epidemic spreading process, a small number of nodes
are randomly chosen to be infected and the other nodes are in
susceptible state. At each time step, the disease first propagates
from every infected node to all its neighbors.
When a neighbor of one infected node is in the susceptible state,
it will be infected with probability
$\lambda(w)=1-(1-\beta)^{w}$, where $w$ is the weight of edge
linking the two nodes and $\beta$ is the unit infection probability for
$w=1$. Obviously, $\lambda(w)$ increases with $w$.
If a susceptible node $i$ has $\Gamma_i$ infected neighbors,
it will be infected with probability $1-\Pi_{j\in \Gamma_i}[1-\lambda(w_{ij})]$,
where $w_{ij}$ is the edge-weight between node $i$ and its infected neighbor $j$.
At the same time step, each infected node can enter into the recovery state with
probability $\gamma$. To be concrete, we set $\gamma=1.0$. Once an
infected node is recovered, it will remain in this state for all
subsequent times.

\section{edge-weight based compartmental approach}
\label{sec:MF_theory}

Two key quantities in the spreading dynamics are the
epidemic threshold and epidemic size~(\emph{i.e.}, final infected
density). We first develop an edge-weight based compartmental
approach to predict these two quantities on the networks with arbitrary degree
and weight distributions. Then, we investigate the effectiveness of the
edge-weight based removal strategy on epidemic spreading through the proposed method.
The time evolution of the epidemic spreading is described by the variables
$S(t)$, $I(t)$ and $R(t)$, which are the densities of the
susceptible, infected, and recovered nodes at time $t$,
respectively.

\subsection{Spreading dynamics} \label{subsec:MF_rate_equation}
In the classical heterogeneous mean-field theory (CHMF), nodes are
classified according to their degrees, which means all the nodes within a
given class are considered to be statistically
equivalent~\cite{Moreno2002,Castellano2006}.
However, apart from the heterogeneity of degree distribution,
the heterogeneity of weights on edges makes the CHMF theory be
hard to accurately describe the spreading dynamics on weighted networks~\cite{Buono2013}.
To solve this question, we develop an edge-weight based compartmental
theory, which is inspired by Refs.~\cite{Volz2008,Volz2011}.

We define $\theta_{w}(t)$ to be the probability that a node $v$ has
not transmitted the infection to a node $u$ along a randomly chosen
edge with weight $w$. Initially, only a few nodes are in the
infected state, which means $\theta_{w}(t)$ is close to unity. A
randomly selected node $u$ is not infected by one of its neighbors
with probability
\begin{equation} \label{theta}
\theta(t)=\sum_{w}g(w)\theta_{w}(t).
\end{equation}
By time $t$, if none of its neighbor has transmitted the infection to node $u$,
it will remain in the susceptible state. Supposing its degree is $k$, it is susceptible at time $t$ with
probability $\theta(t)^{k}$. Thus, the proportion of the susceptible
nodes (\emph{i.e.}, the probability that a randomly selected node is
susceptible) at time $t$ is
\begin{equation} \label{susceptible}
S(t)=\sum_{k=0}p(k)\theta(t)^{k}=G(\theta(t)),
\end{equation}
where $G(x)=\sum_{k}p(k)x^{k}$ is the generating function for degree
distribution.

A neighbor of node $u$ may be in one of susceptible,
infected and recovered states, and thus the probability $\theta_w(t)$
for weight $w$ can be divided into three parts:
\begin{equation} \label{sum_theta}
\theta_w(t)=\xi_w^{S}(t)+
\xi_{w}^{I}(t)+\xi_{w}^{R}(t),
\end{equation}
where $\xi_w^{S}(t)$ ($\xi_w^{I}(t)$ or $\xi_w^{R}(t)$)
is denoted as the probability that a neighbor is in the susceptible
(infected or recovery) state and has not
transmitted the infection to node $u$ through an edge with weight $w$ by time $t$.
Once these three parameters can be derived,
we will get the density of susceptible nodes at time $t$
by substituting them into Eq.~(\ref{theta})
and then into Eq.~(\ref{susceptible}). To this purpose, in the
following, we will focus on how to solve them.

If a neighbor of node $u$ is susceptible, it can not infect the node
$u$, and vice versa. On the uncorrelated networks, one link of node $u$ connects to
a node with degree $k$ with probability $kp(k)/\langle k\rangle$,
where $\langle k\rangle$ is the mean degree of a network~\cite{Caldarelli2003}.
In the mean-field level, the probability that one of its neighbors
is in susceptible state is $\xi_w^{S}(t)=\Sigma_{k}kp(k)\theta(t)^{k-1}/\langle k\rangle$.
Utilizing the generating function for degree distribution $G(x)$,
we have
\begin{equation} \label{xiS}
\xi_{w}^{S}(t)=\frac{G^{\prime}(\theta(t))}{G^{\prime}(1)}.
\end{equation}
According to the spreading
process described in Sec.~\ref{sec:model}, we know that the
growth of $\xi_{w}^{R}(t)$ includes two consecutive events:
firstly, an infected neighbor has not transmitted the infection to node $u$
via their edge with weight $w$, with probability $1-\lambda(w)$;
secondly, the infected neighbor has been recovered, with probability
$\gamma=1.0$. Combining these two events, we have
\begin{equation} \label{xiR}
\frac{d\xi_{w}^{R}(t)}{dt}=(1-\lambda(w))\xi_{w}^{I}(t).
\end{equation}
If this infected neighbor transmits the infection via an edge with weight $w$,
the rate of flow from $\theta_w(t)$ to $1-\theta_{w}(t)$ will be
$\lambda(w)\xi_{w}^{I}(t)$, which means
\begin{equation} \label{theta_Rate}
\frac{d\theta_{w}(t)}{dt}=-\lambda(w)
\xi_{w}^{I}(t),
\end{equation}
and
\begin{equation} \label{theta_Rate1}
\frac{d(1-\theta_{w}(t))}{dt}=\lambda(w)
\xi_{w}^{I}(t).
\end{equation}
By combining Eqs.~(\ref{xiR}) and~(\ref{theta_Rate1}), one obtains
\begin{equation} \label{xiR_C}
\xi_{w}^{R}=\frac{(1-\theta_{w}(t))(1-\lambda(w))}{\lambda(w)}.
\end{equation}
Substituting Eq.~(\ref{xiS}) of $\xi_{w}^{S}(t)$ and
Eq.~(\ref{xiR_C}) of $\xi_{w}^{R}(t)$ into Eq.~({\ref{sum_theta}}),
we yield the following relation
\begin{equation} \label{xiI}
\xi_{w}^{I}(t)=\theta_w(t)-\frac{G^{\prime}(\theta(t))}{G^{\prime}(1)}
-(1-\theta_{w}(t))\frac{1-\lambda(w)}{\lambda(w)}.
\end{equation}
Injecting Eq.~(\ref{xiI}) into Eq.~(\ref{theta_Rate}), we have
\begin{equation} \label{theta_w}
\frac{d\theta_w(t)}{dt}=\lambda(w)\frac{G^{\prime}(\theta(t))}{G^{\prime}(1)}\\
+1-\lambda(w)-\theta_w(t).
\end{equation}
From Eq.~(\ref{theta_w}), the probability $\theta_w(t)$ can be solved.
Substituting the value of $\theta_w(t)$ into Eqs.~(\ref{theta}) and (\ref{susceptible}),
the density associated with each distinct state is given by
\begin{eqnarray}
\label{R}
\frac{dR(t)}{dt} & = & I(t), \\
\label{S}
S(t) & = & G(\theta(t)),\\
\label{I}
I(t) & = & 1-R(t)-S(t).
\end{eqnarray}

According to Eqs.~(\ref{theta_w})-(\ref{I}), one can find that only
 $E\propto O(w_{max})$ equations are required in the edge-weight based compartmental
approach to describe the dynamics of epidemic on networks with
arbitrary degree and weight distributions. By setting $t\rightarrow
\infty$ and $d\theta_w/dt=0$ in Eq.~(\ref{theta_w}), we get
the probability of one edge with weight $w$ which does not propagate
disease in the spreading process
\begin{equation} \label{fina_theta}
\theta_w(\infty)=\lambda(w)\frac{G^{\prime}(\theta(\infty))}{G^{\prime}(1)}\\
+1-\lambda(w).
\end{equation}
Substituting $\theta_w(\infty)$ into Eqs.~(\ref{theta}) and (\ref{susceptible}),
we can figure out the value of $S(\infty)$, and then the epidemic size $R(\infty)$ can be obtained.

Another important issue in epidemic spreading is to
determine the epidemic threshold. Below the epidemic threshold,
the epidemic will die out; otherwise, the epidemic will spread and
become possible. To this end, we summate $\theta_w(\infty)$ for
all possible $w$ in Eq.~(\ref{fina_theta}) and obtain
\begin{equation} \label{fina_theta_2}
\theta(\infty)=\langle\lambda(w)\rangle\frac{G^{\prime}(\theta(\infty))}{G^{\prime}(1)}\\
+1-\langle\lambda(w)\rangle,
\end{equation}
where
\begin{equation} \label{transmission rate}
\langle \lambda(w)\rangle=\sum_{w}g(w)\lambda(w)
\end{equation}
is the mean transmission rate for a randomly selected edge. Indeed, we
can obtain the threshold of epidemic by observing where the
non-trivial solution of Eq.~(\ref{fina_theta_2}) appears,
corresponding to the point at which the right-hand side of the
equation is tangent to the line $y=\theta(\infty)$ at
$\theta(\infty)=1$~\cite{Newman2010}. The condition of the epidemic threshold
is thus given by
\begin{equation} \label{threshold_0}
\langle \lambda_c(w)\rangle=\frac{G^{\prime}(1)}{G^{\prime\prime}(1)}
=\frac{\langle k\rangle}{\langle k^2\rangle-\langle k\rangle}.
\end{equation}
Further solving the above equation, one can get the epidemic threshold
\begin{equation} \label{threshold}
\beta_c=1-F^{-1}(z),
\end{equation}
where $F(x)=\sum_w g(w)x^w$ is the generating function for weight distribution,
$F^{-1}(z)$ is the inverse function of $F(x)$, and $z=1-\langle k\rangle/(\langle
k^2\rangle-\langle k\rangle)$.

From Eq.~({\ref{threshold}}), we see that the epidemic threshold is
closely related to the degree and weight distributions.
For a given weight distribution, the stronger heterogeneity
of degree distribution with the larger value of $z$
results in the smaller value of $\beta_c$,
as $F^{-1}(z)$ is a monotone increasing function.
By contrast, increasing the heterogeneity of weight distribution
can enhance the threshold of epidemic outbreak when the degree
distribution is fixed, since $F^{-1}(z)$ decreases with
the heterogeneity of weight distribution at $0<z<1$\textbf.
If the weight on every edge equals to unity,
the epidemic threshold will return to the result obtained on
unweighted networks~\cite{Newman2002}.

\subsection{Effectiveness of edge removal} \label{subsec:threshold}

To prevent an epidemic in time, different strategies to immunize nodes or edges of a network have
been widely studied~\cite{Pastor-Satorras2002_2,Cohen2003}.
A successful immunization strategy must be able to accurately identify and
immunize the influential nodes or edges in the process of epidemic spreading,
which can significantly enhance the epidemic threshold and reduce the epidemic size~\cite{Chen2008}.
In weighted networks, the edge weight reflects the relative importance of the connections between nodes,
and edges with high weights may play a more significant role in the spreading process~\cite{Li2013}.
In the ideal case with full knowledge of all edge weights,
removing the edges with high weights can prevent epidemic spreading more efficiently.
But in many realistic cases, this complete information is not available,
and only partial knowledge exists~\cite{Gong2013,Yang2012_2,Gallos2005}.
Therefore, we here focus on a general edge-weight based removal model with
the family of function~\cite{Gallos2005,Huang2011}
\begin{equation} \label{chi}
\Phi(w)=\frac{w^{\alpha}}{\sum_{i=1}^{M}w_{i}^{\alpha}}, -\infty<\alpha<+\infty,
\end{equation}
where a value $\Phi(w)$ is assigned to each edge, which stands for the probability that an edge
with weight $w$ is removed, $M$ is the total number of edges,
and $\alpha$ is an exponent of preferential removal.
For $\alpha=0.0$, we have $\Phi(w)=1/M$, which means every
edge has the same removal probability. The case of
$\alpha\rightarrow+\infty$ represents that the strategy
is to remove edges in a descending order (\emph{i.e.},
from edges with high weights to edges with low weights). For
$\alpha\rightarrow-\infty$, the opposite case happens.
After an edge removal strategy is executed, where a fraction $1-f$ of edges are
removed from the original network according to Eq.~(\ref{chi}),
we initiate an infection on the residual network.

To obtain the epidemic threshold and epidemic size, we first figure out
the degree and weight distributions of the residual network,
and then solve these two key quantities
through the edge-weight based compartmental theory in Sec. III. A.
In our network model, degree and weight distributions are respectively
independent, which means a fraction of $1-f$ edges are randomly
removed in the edge removal strategies with different values of $\alpha$.
Equivalently, the residual network can be gotten
by randomly occupying a fraction of $f$ edges in the original network.
Using the percolation theory, the degree distribution of the
residual network is given by~\cite{Newman2002,Newman2010}
\begin{equation} \label{after_degree}
p_{f}(k)=\sum_{m=k}p(m) {m\choose k}f^{k}(1-f)^{m-k}.
\end{equation}

Letting $A_{f}(w)$ be the number of edges with weight $w$ and $g_{f}(w)$
be the residual weight distribution in the residual weighted network
with the remaining fraction $f$ of edges,
we have the residual weight distribution as
\begin{equation} \label{after_weight}
g_{f}(w)=\frac{A_{f}(w)}{fM}.
\end{equation}
When one additional edge is removed by implementing the edge weight based removal strategy
as Eq.~(\ref{chi}),
$A_{f}(w)$ becomes
\begin{equation} \label{another_edge}
A_{(f-\frac{1}{M})}(w)=A_{f}(w)-\frac{g_{f}(w)w^{\alpha}}{\langle w^{\alpha}(f)\rangle},
\end{equation}
where $\langle w^{\alpha}(f)\rangle=\sum_{w}g_{f}(w)w^{\alpha}$. In the
thermodynamic limit $M\rightarrow\infty$, Eq.~(\ref{another_edge}) can be presented in terms
of derivative of $A_{f}(k)$ with respect to $f$,
\begin{equation} \label{diff}
\frac{dA_{f}(w)}{df}=M\frac{g_{f}(w)w^{\alpha}}{\langle w^{\alpha}(f)\rangle}.
\end{equation}
Differentiating Eq.~(\ref{after_weight}) with respect to $f$ and
substituting it into Eq.~(\ref{diff}), we obtain
\begin{equation} \label{diff_1}
-f\frac{dg_{f}(w)}{df}=g_{f}(w)-\frac{g_{f}(w)w^{\alpha}}{\langle w^{\alpha}(f)\rangle}.
\end{equation}
To solve Eq.~(\ref{diff_1}), we define a function
$H_{\alpha}(t)=\sum_{w}g(w)t^{w^{\alpha}}$, and let
$t=H_{\alpha}^{-1}(f)$. We find by direct differentiation
that~\cite{Shao2009,Huang2011}
\begin{equation} \label{after_weight_1}
g_{f}(w)=g(w)\frac{t^{w^{\alpha}}}{H_{\alpha}(t)}=\frac{1}{f}g(w)t^{w^{\alpha}},
\end{equation}
and
\begin{equation} \label{mean_w}
\langle w^{\alpha}(f)\rangle=\frac{tH_{\alpha}^{\prime}(t)}{H_{\alpha}(t)}.
\end{equation}
From Eqs.~(\ref{after_degree}) and~(\ref{after_weight_1}), we can get the
degree and weight distributions of the residual weighted network, respectively.
Substituting them into Eqs.~(\ref{R})-(\ref{fina_theta}) and Eq.~(\ref{threshold}),
we can obtain the epidemic size and outbreak condition on the
residual network, respectively.

\section{Numerical simulations}
\label{sec:UCMN_simulation}

In simulations, the size of networks, the mean degree and the
mean edge weight are set to be $N=10^4$, $\langle k\rangle=10$
and $\langle w\rangle=8$, respectively. Without of lose generality,
we consider the networks with degree distribution
$p(k)\sim k^{-\gamma_{D}}$ and weight distribution $g(w)\sim
w^{-\gamma_{W}}$ to verify the theoretical approach, where $\gamma_{D}$ and $\gamma_{W}$ represent
degree and weight exponents, respectively. The smaller
values of the exponents, the more heterogeneous of the
distributions~\cite{Newman2001_2}. The maximum degree and weight are
set to be $k_{max}\sim \sqrt{N}$ ~\cite{Clauset2009} and
$w_{max}\sim N^{1/(\gamma_W-1)}$~\cite{Yang2012}, respectively.
To initiate an infection process, we randomly choose five infected nodes as seeds,
while the other nodes are in the susceptible state.

We employ the susceptibility measure \cite{Ferreira2012,Wang2014} $\chi$
to numerically determine the epidemic threshold
\begin{equation}\label{susceptibility}
\chi=N\frac{\langle r^{2}\rangle-\langle r\rangle^{2}}{\langle r\rangle},
\end{equation}
where $r$ denotes the epidemic size $R$. To obtain a
reliable value of $\chi$, we use at least $2\times10^3$ independent
dynamic realizations on a fixed weighted network to calculate the
average value of $\chi$ for each value of unit infection probability $\beta$.
Susceptibility $\chi$ exhibits a maximum value at $\beta_c$,
which is the threshold value of the epidemic spreading process.
The simulations are further implemented  by using $100$ different network realizations
to obtain the mean threshold $\beta_{c}$. The identical
simulation setting is used for all subsequent numerical results,
unless otherwise specified.

\begin{figure}
\begin{center}
\epsfig{file=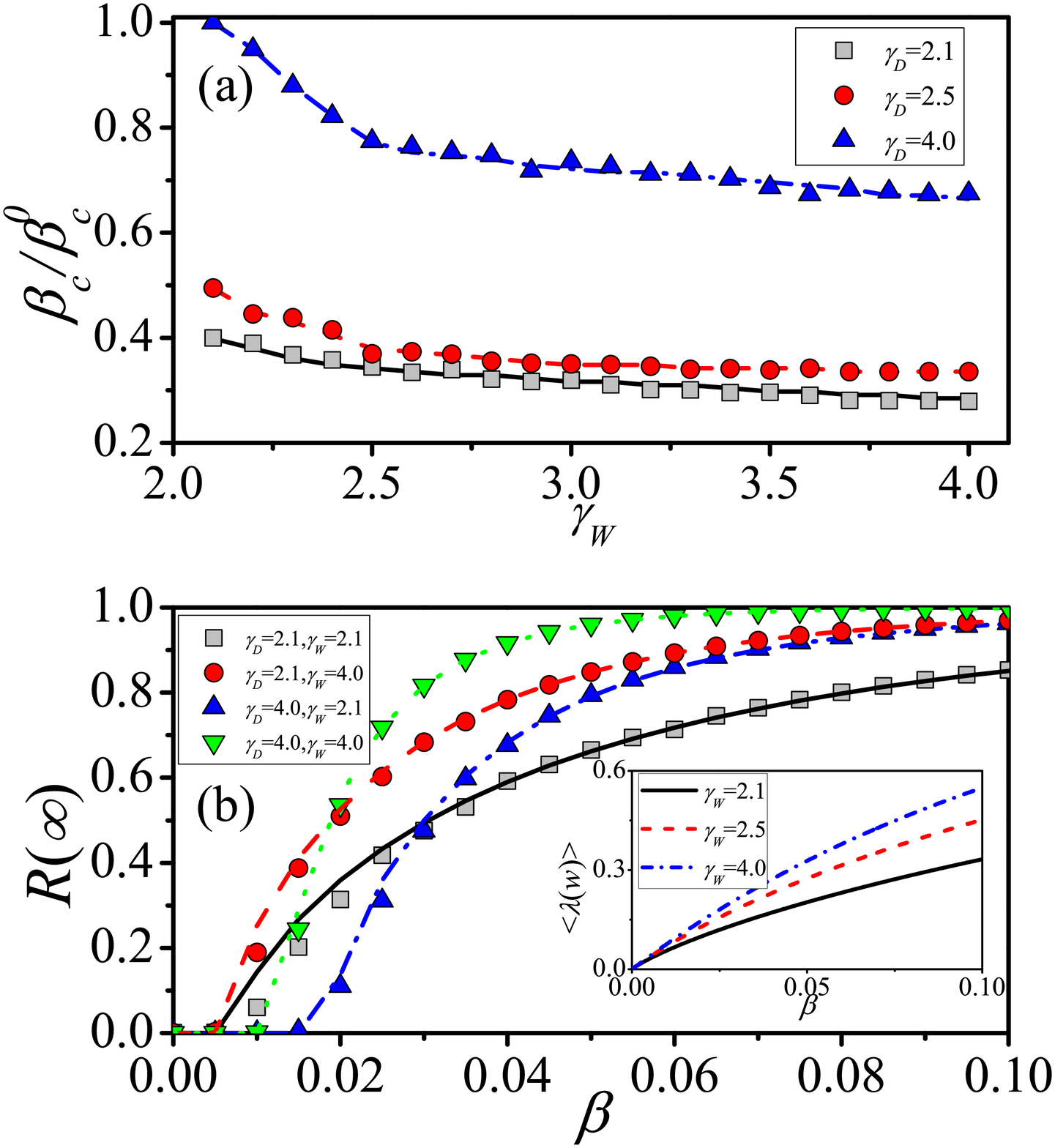,width=1\linewidth}
\caption{(Color online) The influence of degree and weight
distributions on the epidemic threshold and epidemic size. (a) The
relative epidemic threshold $\beta_{c}/\beta_{c}^{0}$ as a function of weight
exponent $\gamma_{W}$ for degree exponents $\gamma_{D}=2.1$~(gray
squares), $\gamma_{D}=2.5$~(red circles) and $\gamma_{D}=4.0$~(blue
up triangles), where $\beta_{c}^{0}\approx0.016$ is the theoretical
threshold at $\gamma_{D}=4.0, \gamma_{W}=2.1$.  Black solid, red
dashed and blue dot-dashed lines are the numerical solutions from
Eq.~(\ref{threshold}). (b) Epidemic size $R(\infty)$ versus
unit infection probability $\beta$ for $\gamma_{D}=2.1, \gamma_{W}=2.1$~(gray squares),
$\gamma_{D}=2.1, \gamma_{W}=4.0$~(red circles), $\gamma_{D}=4.0,
\gamma_{W}=2.1$ (blue up triangles) and $\gamma_{D}=4.0,
\gamma_{W}=4.0$~ (green down triangles). Black solid, red dashed,
blue dot-dashed and green dot lines are the numerical solutions from
Eqs.~(\ref{R})-(\ref{fina_theta}). The inset of (b) shows the
numerical solutions of $\langle\lambda(w)\rangle$ from Eq.~(\ref{transmission rate})
as function of  $\beta$ for three different values of
$\gamma_{W}$~(\emph{i.e.,} 2.1, 2.5, and 4.0), corresponding to the black solid,
red dashed and blue dot-dashed lines.} \label{fig:spreading_T}
\end{center}
\end{figure}

We first investigate the influence of degree distribution on the epidemic dynamics.
Fig.~\ref{fig:spreading_T} (a) shows that the epidemic threshold $\beta_c$ decreases with
the heterogeneity of degree distribution~(\emph{i.e.}, the smaller value of
$\gamma_{D}$) on the weighted networks, which is consistent with the epidemic outbreak on the unweighted
networks~\cite{Pastor-Satorras2001}. This reason stems from the
existence of more hub nodes on strong heterogeneous networks.
However, the effect of degree distribution on the epidemic size $R(\infty)$ is more
complex. As shown in Fig.~\ref{fig:spreading_T} (b),
increasing the heterogeneity of degree distribution can promote the epidemic size
at small $\beta$ while suppress the epidemic size at large $\beta$.
For instance, at a fixed value of $\gamma_{W}=2.1$, the epidemic size $R(\infty)$
for $\gamma_D=2.1$ is greater than that for $\gamma_D=4.0$ when $\beta\leq0.03$ (\emph{i.e.,} promotion region),
while the situation is exactly opposite when $\beta>0.03$ (\emph{i.e.,} suppression region).
This result can be qualitatively explained as follows: epidemic
propagates on complex networks following a hierarchical way. That is
to say, the hubs with large degrees are more likely to
become infected at the early times of epidemic spreading~\cite{Barthelemy2004}.
For a small value of $\beta$, the existence of hubs
makes the epidemic spread more easily. But for a large value of $\beta$,
more nodes with small degrees in more heterogeneous networks have a small infection probability,
which results in the lower $R(\infty)$.

The effects of heterogeneity of weight distribution on the epidemic threshold as well
as the epidemic size are also given in Fig.~\ref{fig:spreading_T}. One
can find that, when the degree distribution (\emph{i.e.}, the
value of $\gamma_{D}$) is fixed, increasing the heterogeneity of weight
distribution (\emph{i.e.}, decreasing the value of $\gamma_{W}$) not
only enhances the epidemic threshold $\beta_c$ [see Fig.~\ref{fig:spreading_T}(a)]
but also reduces the epidemic size $R(\infty)$ [see
Fig.~\ref{fig:spreading_T}(b)]. This phenomenon can be explained as
follows: when the average weight $\langle w\rangle$ is fixed,
the small value of $\gamma_{W}$ causes most edges possessing
lower weights and infection probabilities, leading to the fact that
the mean transmission rate $\langle\lambda(w)\rangle$
for a randomly selected edge is smaller on the networks with more heterogeneous weight distribution
[see Eq.~(\ref{transmission rate}) and its numerical solutions in the inset of Fig.~\ref{fig:spreading_T}(b)].
In addition, from Fig.~\ref{fig:spreading_T}(a) one can see that the epidemic
threshold $\beta_c$ increases more remarkably when $\gamma_W\leq2.5$ due to
the strong heterogeneity of the weight distribution. On these networks with
strong heterogeneous degree and weight distributions, the developed edge-weight based approach
can still accurately reproduce the simulated $\beta_c$ and $R(\infty)$ in Fig.~\ref{fig:spreading_T}.

\begin{figure}
\begin{center}
\epsfig{file=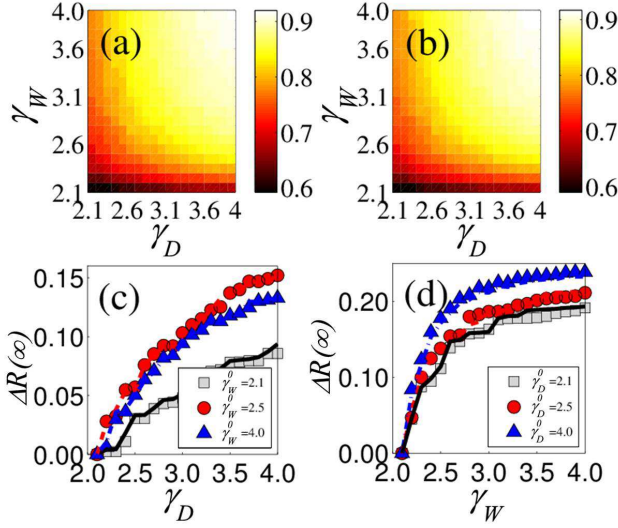,width=1\linewidth}
\caption{(Color online) Epidemic size changes with the
degree and weight exponents. Color-coded values of epidemic size
from numerical simulations (a) and theoretical solutions (b) are shown on the $\gamma_D, \gamma_W$ plane.
(c) The increment of epidemic size $\Delta R(\gamma_D,\gamma_W^{0},\infty)$
as a function of degree exponent $\gamma_D$ at weight exponents $\gamma_W^0=2.1$~(gray
squares), $\gamma_W^0=2.5$~(red circles) and $\gamma_W^0=4.0$~(blue
up triangles). (d) $\Delta R(\gamma_D^{0},\gamma_W,\infty)$
as a function of $\gamma_W$ at $\gamma_D^0=2.1$~(gray squares),
$\gamma_D^0=2.5$~(red circles) and $\gamma_D^0=4.0$~(blue up triangles).
The values in (b) and the lines in (c) and (d) are the numerical solutions
of Eqs.~(\ref{R})-(\ref{fina_theta}) in the limit $t\rightarrow\infty$.
The unit infection probability is set to $\beta=0.04$.}
\label{fig2}
\end{center}
\end{figure}

To further investigate the impacts of the two heterogeneous distributions on
the epidemic  size, $R(\infty)$ as a function of the exponents
$\gamma_D$ and $\gamma_W$ is shown in Fig.~\ref{fig2},
where the unit infection probability is set to $\beta=0.04$
in the suppression region for ensuring the outbreak of epidemic.
From Figs.~\ref{fig2} (a) and (b), we see that the weaker heterogeneity
of degree and weight distributions can lead to the higher epidemic size,
that is, $R(\infty)$ increases with the growth of $\gamma_D$ and $\gamma_W$.
We also show the increase of $R(\infty)$ with $\gamma_D$ ($\gamma_W$)
at a fixed value of $\gamma_W^{0}$ ($\gamma_D^{0}$) in Fig.~\ref{fig2} (c) [Fig.~\ref{fig2} (d)].
Defining the increment of $R(\infty)$ as $\Delta R(\gamma_D,\gamma_W,\infty)=R(\gamma_D,\gamma_W,\infty)-R(\gamma_D^{0},\gamma_W^{0},\infty)$,
where $R(\gamma_D,\gamma_W,\infty)$ is the epidemic size for a $(\gamma_D,\gamma_W)$ pair,
we can investigate the impact of one parameter on the value of $\Delta R(\gamma_D,\gamma_W,\infty)$ by fixing the
other parameter. For example, by setting $\gamma_W^{0}=2.1$ we can
look into how $\Delta R(\gamma_D,\gamma_W^{0},\infty)$ changes with $\gamma_D$.
We note that the increment of $R(\infty)$ tends to be more evident
for the more homogeneous weight and degree distributions
(\emph{i.e.,} greater $\gamma_W$ and $\gamma_D$),
which results from the greater mean transmission rate $\langle\lambda(w)\rangle$
and fewer nodes with small degrees having a small infection probability, respectively.
From Fig.~\ref{fig2}, we see that the theoretical predictions
are in good agreement with the simulated epidemic size,
no matter how heterogeneous the degree and weight distributions are.

\begin{figure}
\begin{center}
\epsfig{file=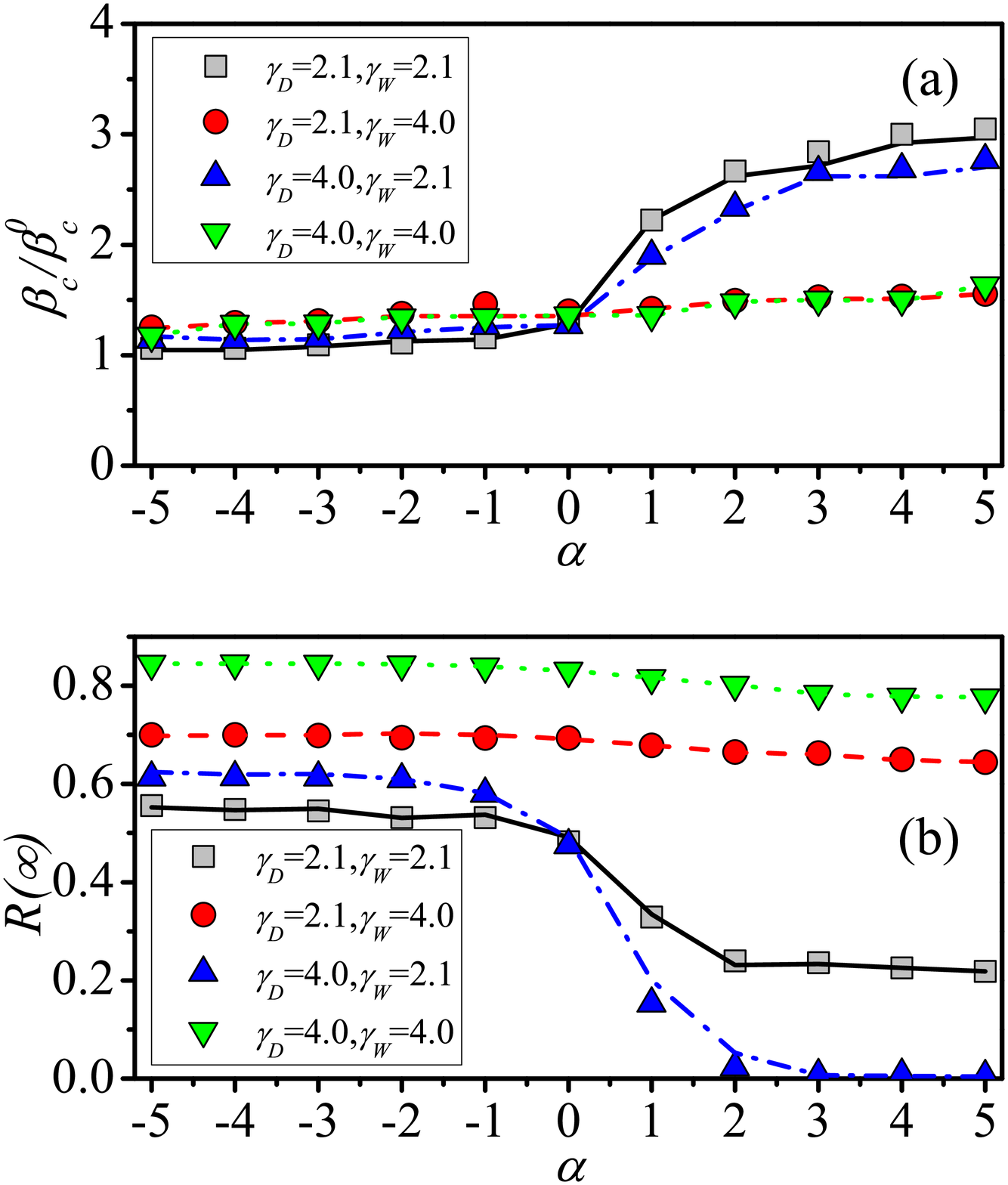,width=1\linewidth}
\caption{(Color online) The control effectiveness of the edge-weight
based removal strategy with different bias.
The relative epidemic threshold $\beta_c/\beta_c^0$ (a)
and epidemic size $R(\infty)$ (b) as a function of parameter
$\alpha$ on different networks with tunable parameters
$\gamma_D=2.1,\gamma_W=2.1$~(gray squares),  $\gamma_D=2.1,\gamma_W=4.0$
~(red circles), $\gamma_D=4.0,\gamma_W=2.1$~(blue up triangles) and
$\gamma_D=4.0,\gamma_W=4.0$~(green down triangles).
Black solid, red dashed, blue dot-dashed and green dot lines are the
analytical predictions from Eq.~(\ref{threshold}) for the relative threshold
and Eqs.~(\ref{R})-(\ref{fina_theta}) for the epidemic size,
with the degree and weight distributions according to Eqs.~(\ref{after_degree})
and (\ref{after_weight_1}), respectively. In subfigure (a), $\beta_{c}^{0}$
is the theoretical threshold on the original network.
}
\label{fig3}
\end{center}
\end{figure}

In the following, we check the effectiveness of the edge-weight based
removal strategy on controlling epidemics. Fig.~\ref{fig3}
reports the epidemic threshold and epidemic size as a function of
the tunable parameter $\alpha$ when a fraction $1-f=0.2$ of edges
are removed according to Eq.~(\ref{chi}). For $\alpha>0.0$,
preferentially removing edges with high weights can partly restrain the spread of epidemic (\emph{i.e.},
enhance the epidemic threshold and reduce the epidemic size).
The reason of this phenomenon is that the removal of strong ties (\emph{i.e.},
edges with high weights) can reduce the value of $\langle\lambda(w)\rangle$ more effectively
than that of weak ties (\emph{i.e.}, edges with low weights).
So the control effect for $\alpha\leq0.0$ is negligible,
as the removal of edges is concentrated on the weak ties.
In addition, one can see that the more heterogenous
the weight distribution is (\emph{i.e.}, the smaller value of $\gamma_W$),
the better effectiveness the edge-weight based removal strategy plays when $\alpha>0.0$.
For $\gamma_W=2.1$ in Fig.~\ref{fig3} (a), the edge removal strategy
with large $\alpha$ makes the epidemic threshold $\beta_c$ increases
by two or three times, that is the relative threshold $\beta_c/\beta_c^0\approx2.5$,
where $\beta_c$ and $\beta_c^0$ are respectively the
epidemic thresholds for the residual and original networks.
Fig.~\ref{fig3} (b) also shows that the epidemic can
almost be eliminated on the networks with $\gamma_D=4.0$ and
$\gamma_W=2.1$ when $\alpha\geq2.0$ [see the blue up triangles in
Fig.~\ref{fig3} (b)], because the epidemic threshold $\beta_c$ is
close to $0.04$ after removing many strong ties [see Fig.~\ref{fig3} (a)].

\begin{figure}
\begin{center}
\epsfig{file=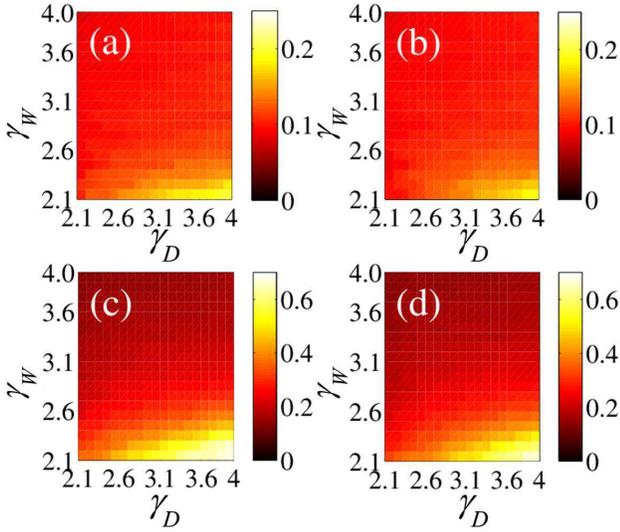,width=1\linewidth}
\caption{(Color online) The decrement of epidemic size $\triangle R'(\infty)$
as a function of $\gamma_D$ and $\gamma_W$ for different values of $\alpha$.
(a) and (c) represent respectively the numerical simulations for $\alpha=0.0$ and $\alpha=5.0$,
and the theoretical predications for $\alpha=0.0$ and $\alpha=5.0$
are shown in (b) and (d), respectively. The unit infection probability is $\beta=0.04$.
}
\label{fig4}
\end{center}
\end{figure}

We further address the performance of this removal strategy
on reducing the epidemic size for $\beta=0.04$ in Fig.~\ref{fig4}.
The decrement of epidemic size is defined as $\Delta
R^\prime(\gamma_D,\gamma_W,\infty)=R_0(\gamma_D,\gamma_W,\infty)-R(\gamma_D,\gamma_W,\infty)$
, where $R_0(\gamma_D,\gamma_W,\infty)$ and
$R(\gamma_D,\gamma_W,\infty)$ are the epidemic sizes on the
original network and residual network, respectively.
Figs.~\ref{fig4} (a) and (b) reveal that,
$\Delta R^\prime(\gamma_D,\gamma_W,\infty)$ is small for $\alpha=0.0$,
as the value of $\langle\lambda(w)\rangle$ is little changed
and the mean degree of the residual network $\langle k\rangle$ is slightly smaller,
when some edges are randomly removed. We also see that the
$\Delta R^\prime(\gamma_D,\gamma_W,\infty)$ is smaller
for the case of small $\gamma_D$ when $\gamma_W$ is small,
as the existence of hubs makes the epidemic spreading
near the threshold has a better robustness against random edge failures~\cite{Newman2010}.
In Figs.~\ref{fig4} (c) and (d), the edge-weight based
removal strategy with $\alpha=5.0$ is more effective,
especially for the networks with strong heterogeneous weight distribution,
which results from the $\langle\lambda(w)\rangle$ decreasing faster.

\begin{figure}
\begin{center}
\epsfig{file=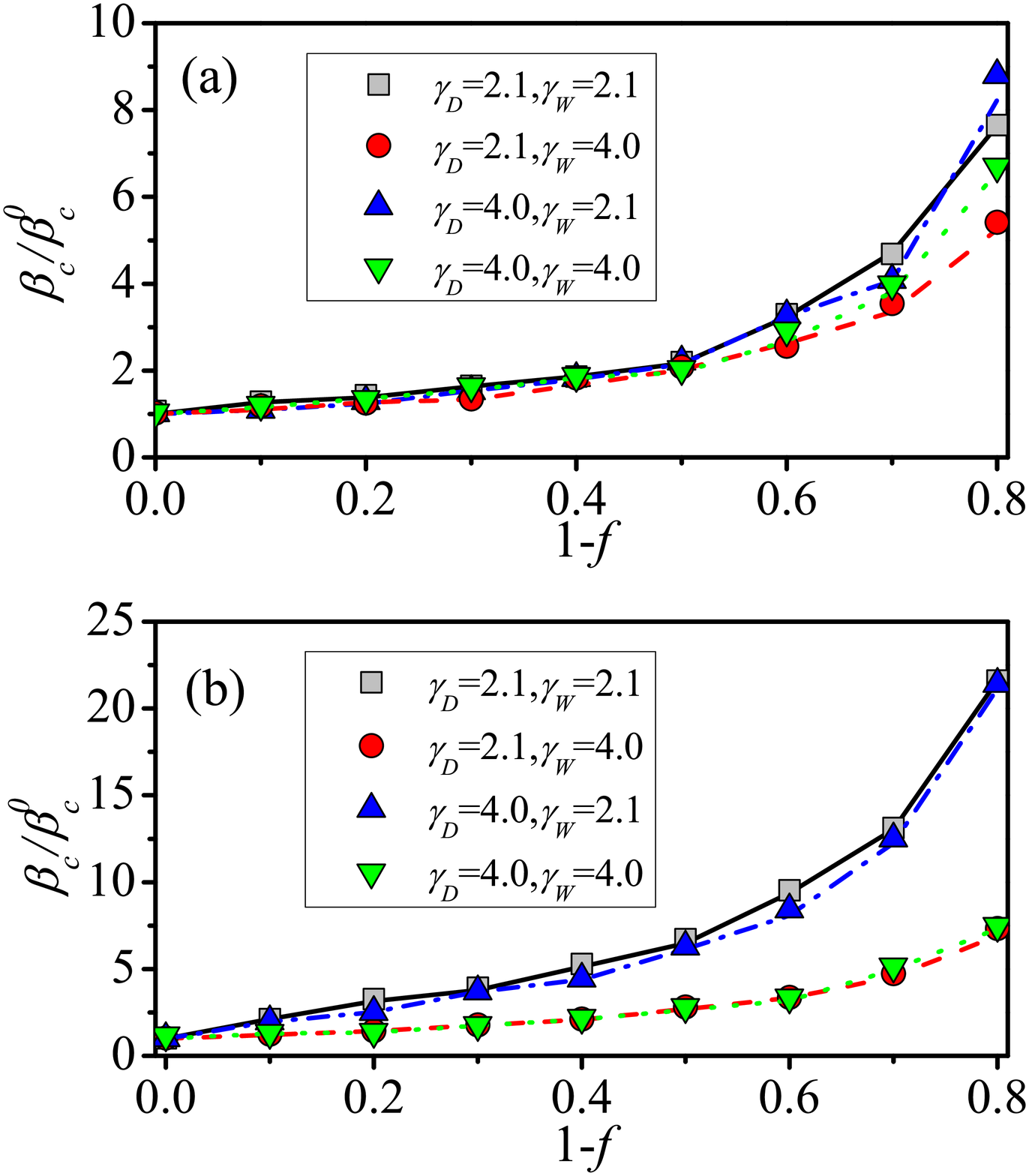,width=1\linewidth}
\caption{(Color online) The relative epidemic threshold
as a function of removal proportion $1-f$ for (a) $\alpha=0.0$
and (b) $\alpha=5.0$ on different networks, including the
parameters $\gamma_D=2.1,\gamma_W=2.1$~(gray squares),  $\gamma_D=2.1,\gamma_W=4.0$
~(red circles), $\gamma_D=4.0,\gamma_W=2.1$~(blue up triangles) and
$\gamma_D=4.0,\gamma_W=4.0$~(green down triangles).
Black solid, red dashed, blue dot-dashed and
green dot lines are the analytical predictions from Eq.~(\ref{threshold})
with the degree and weight distributions according to Eqs.~(\ref{after_degree})
and (\ref{after_weight_1}), respectively. And the parameter
$\beta_{c}^{0}$ is the theoretical threshold on the original network.
}
\label{fig5}
\end{center}
\end{figure}

Moreover, we study the influence of edge removal
proportion on the epidemic threshold for $\alpha=0.0$ [see
Fig.~\ref{fig5} (a)] and $\alpha=5.0$ [see Fig.~\ref{fig5} (b)].
For comparison, we define the relative threshold $\beta_c/\beta_c^{0}$
as the ratio of the threshold $\beta_c$ on the residual network to
the threshold $\beta_c^{0}$ on the original network.
As is expected, increasing the immunization proportion $1-f$ results in
the increase of the relative threshold $\beta_c/\beta_c^{0}$.
We also note that the targeted edge removal strategy with $\alpha=5.0$
presents a much better performance on the weighted networks
with small $\gamma_W$, \emph{e.g.}, $\beta_c/\beta_c^{0}(\gamma_W=4.0)<\beta_c/\beta_c^{0}(\gamma_W=2.1)$ in Fig.~\ref{fig5} (b).
What's more, the simulated results are well fitted by the
theoretical predictions presented in Sec.~\ref{subsec:threshold}.

\section{Conclusions} \label{sec:conclusion}
In sum, in this paper, we developed an edge-weight based approach
to describe the spread of epidemic on the networks with heterogenous
degree distribution as well as heterogenous weight distribution.
Our findings indicate that the predictions from such a method can be
in good agreement with the simulated epidemic threshold and epidemic size.
Combing the numerical simulations and the theoretical analysis,
we found that the \emph{strong} heterogeneity of degree distribution
and the \emph{weak} heterogeneity of weight distribution can both
make networks be more fragile to the outbreak of epidemics.
Unlike the effect of weak heterogeneity of weight distribution
which always promotes epidemic spreading,
the effects of the heterogeneity of degree distribution on
the epidemic size can be divided into two distinct regions:
the strong heterogeneity of degree distribution promotes the epidemic size when
the unit infection probability is small, on the contrary,
the strong heterogeneity suppresses the epidemic size
at a large unit infection probability.
Thus, for a large value of unit infection probability the epidemic spreading
will be mostly promoted once both the degree distribution and the weight
distribution are more homogenous. Moreover, we proposed an edge-weight based
removal strategy and investigated the effectiveness of this strategy
on epidemic control. Generally speaking, removing edges with high
weights is more effective to suppress epidemic spreading
on the networks with strong heterogeneous weight distribution,
especially for the networks having more homogeneous degree distribution
near the epidemic threshold.

We here provides a more accurate theoretical framework
to solve the epidemic spreading on complex networks
with general degree and weight distributions, which
could be applied to other analogous dynamical processes such as
information diffusion and cascading failure. Besides,
how to develop an analytic method for being suitable for the case of
the correlation between nodes' degrees and edge weights existing
still needs to think deeply. This work helps to understand the spreading dynamics
on heterogeneous weighted networks in depth and would stimulate
further works in designing better immunization strategies.

\acknowledgments

This work was partially supported by National Natural Science Foundation of
China (Grant Nos.~11105025, 11135001, 91324002, 11331009), China Postdoctoral Science Special Foundation
(Grant No. 2012T50711), the Program of Outstanding
Ph. D. Candidate in Academic Research by UESTC (Grand No. YXBSZC20131065).
Y. Do was supported by Basic Science Research Program through the National
Research Foundation of Korea (NRF) funded by the Ministry of Education,
Science and Technology (NRF-2013R1A1A2010067).

\end{document}